\documentclass[12pt,letter]{article}
\pdfoutput=1
\usepackage{graphicx, epsfig, color,cite}
\usepackage{amsmath}
\usepackage{amssymb}
\usepackage{float}
\usepackage{subfig}
\usepackage{hyperref}

\def\to{\rightarrow}

\def\bi{\begin{itemize}}
\def\ei{\end{itemize}}

\def\tst{\tilde t}

\def\tg{\tilde g}

\def\alt{\lesssim}
\def\agt{\gtrsim}
\def\be{\begin{equation}}  
\def\ee{\end{equation}}  
\def\bea{\begin{eqnarray}}  
\def\eea{\end{eqnarray}}

\begin{document}
\begin{titlepage}
\begin{flushright}
OU-HEP-241225
\end{flushright}

\vspace{0.5cm}
\begin{center}
  {\Large \bf Implications of Higgs mass\\ for hidden sector SUSY breaking}\\
\vspace{1.2cm} \renewcommand{\thefootnote}{\fnsymbol{footnote}}
{\large Howard Baer$^{1}$\footnote[1]{Email: baer@ou.edu },
Vernon Barger$^2$\footnote[2]{Email: barger@pheno.wisc.edu},
Jessica Bolich$^1$\footnote[3]{Email: Jessica.R.Bolich-1@ou.edu} and
Kairui Zhang$^1$\footnote[5]{Email: kzhang25@ou.edu}
}\\ 
\vspace{1.2cm} \renewcommand{\thefootnote}{\arabic{footnote}}
{\it 
$^1$Homer L. Dodge Department of Physics and Astronomy,
University of Oklahoma, Norman, OK 73019, USA \\[3pt]
}
{\it 
$^2$Department of Physics,
University of Wisconsin, Madison, WI 53706 USA \\[3pt]
}
%{\it 
%$^3$Department of Physics,
%University of Nebraska, Lincoln, NE 68588 USA \\[3pt]
%}
%{\it 
%$^3$Department of Physics and Astronomy,
%University of Hawaii, Honolulu, HI 53706 USA \\[3pt]
%}

\end{center}

\vspace{0.5cm}
\begin{abstract}
\noindent

Hidden sector SUSY breaking where charged hidden sector fields obtain SUSY
breaking vevs once seemed common in dynamical SUSY breaking (DSB).
In such a case, scalars can obtain large masses but gauginos and $A$-terms
gain loop-suppressed anomaly-mediated contributions which may be smaller by
factors of $1/16\pi^2\sim 1/160$. This situation leads to models such
as PeV or mini-split supersymmetry with $m_{scalars}\sim 160 m_{gaugino}$.
In order to generate a light Higgs mass $m_h\simeq 125$ GeV, the scalar
mass terms are required in the 10-100 TeV range, leading to large, unnatural
contributions to the weak scale.
Alternatively, in gravity mediation with singlet hidden sector fields,
then $m_{scalars}\sim m_{gauginos}\sim A$-terms and the large $A$-terms
lift $m_h\to 125$ GeV even for natural values of $m_{\tst_{1}}\sim 1-3$ TeV.
Requiring naturalness, which seems probabilistically preferred
by the string landscape,
then the measured Higgs mass seems to favor singlets in the hidden sector,
which can be common in metastable and retrofitted DSB models.

\end{abstract}
\end{titlepage}
%\pacs{12.60.-i, 95.35.+d, 14.80.Ly, 11.30.Pb}
%12.60.-i   Models beyond the standard model
%95.35.+d   Dark matter

\section{Introduction}
\label{sec:intro}

Weak scale supersymmetry (SUSY)\cite{Drees:2004jm,Baer:2006rs,Dreiner:2023yus} is highly motivated in that it provides
a 't Hooft technically natural\cite{tHooft:1979rat} solution to the Big Hierarchy
problem\cite{Witten:1981nf,Kaul:1981wp}.
In this case, the weak scale $m_{weak}\sim m_{W,Z,h}\simeq 100$ GeV
with $m_{weak}\sim m_{soft} \sim m_{hidden}^2/m_P$ is natural in a theory where
only the Planck mass $m_P$ appears (here, $m_{hidden}\sim 10^{11}$ GeV
is the scale of hidden sector SUSY breaking) in that the model becomes more
(super)symmetric when $m_{hidden}\to 0$.
This situation maintains even for $m_{hidden}\gg 10^{11}$ GeV where an
apparent Little Hierarchy problem (LHP) ensues.

The LHP seems to have emerged in the wake of recent LHC results\cite{Canepa:2019hph,ATLAS:2024lda} which
apparently require $m_{soft}\gg m_{weak}$ and is exemplified by recent
sparticle search limits on gluino ($m_{\tg}\agt 2.3$ TeV\cite{ATLAS:2019vcq,ATLAS:2022ihe,CMS:2019zmd}) and top-squark
($m_{\tst_1}\agt 1.2$ TeV\cite{ATLAS:2020xzu,CMS:2021beq}) masses within the context of simplified SUSY models.
Thus, the LHP is concerned with a smaller hierarchy mismatch wherein
$m_{soft}\sim 1-10$ TeV is $\gg m_{weak}\sim 0.1$ TeV.
This may lead to a violation of {\it practical naturalness:}\cite{Baer:2015rja,Baer:2023cvi}
\begin{quotation}
An observable is {\it practically natural} when all
{\it independent} contributions to the observable are comparable to or
less than the observable in question.
\end{quotation}
Practical naturalness is the form of naturalness which has been used for
instance by Gaillard and Lee\cite{Gaillard:1974hs} to predict the range
of the charm quark mass shortly before it was discovered.

With regards to SUSY as exemplified by the Minimal Supersymmetric Standard Model (MSSM)\cite{Dimopoulos:1981zb}, minimization of the scalar (Higgs) potential allows one
to relate the measured value of the weak scale to the weak scale soft SUSY breaking terms:
\be
m_Z^2/2=\frac{m_{H_d}^2+\Sigma_d^d-(m_{H_u}^2+\Sigma_u^u)\tan^2\beta}{\tan^2\beta -1}-\mu^2\simeq-m_{H_u}^2-\mu^2-\Sigma_u^u(\tst_{1,2}) .
\label{eq:mzs}
\ee
where $m_{H_u}$ and $m_{H_d}$ are soft breaking SUSY Higgs mass terms,
$\mu$ is the SUSY-conserving Higgs/higgsino mass scale and the
$\Sigma_{u,d}^{u,d}$ contain an assortment of loop corrections
that go as $\Sigma_{u,d}^{u,d}\sim \lambda^2 m_{sparticle}^2/16\pi^2$
where $\lambda$ is some associated gauge or Yukawa coupling.
The $\tan\beta\equiv v_u/v_d$ is the usual ratio of Higgs field
vacuum expectation values (vevs).
A numerical measure of practical naturalness comes from the
{\it electroweak} finetuning measure defined as\cite{Baer:2012up}
\be
\Delta_{EW}\equiv max_i|maximal\ term\ on\ RHS\ of\ Eq.\ \ref{eq:mzs} |/(m_Z^2/2) .
\label{eq:dew}
\ee
A value $\Delta_{EW}<30$ requires all terms on the right-hand-side (RHS) of
Eq. \ref{eq:mzs} to lie within a factor 4 of $m_Z$. 
Eq. \ref{eq:dew} is the most conservative measure of naturalness and
is non-optional.
It depends only on the weak scale particle/sparticle mass
spectra and not on the assumed model which generates it.\footnote{Alternative
  measures such as $\Delta_{p_i}$ expand $m_{H_u}^2(weak)$ in terms of various
  high scale soft SUSY breaking parameters which parameterize our ignorance of
  SUSY breaking.
  These parameters are expected to be all correlated in more encompassing
  SUSY theories\cite{Baer:2013gva} leading to overestimates in $\Delta_{p_i}$ by factors of 10-1000
  as compared to $\Delta_{EW}$\cite{Baer:2023cvi}. An alternative measure $\Delta_{HS}\equiv \delta m_{H_u}^2/m_{H_u}^2$ ignores the fact that $\delta m_{H_u}^2$ and $m_{H_u}^2(\Lambda )$
  are not independent\cite{Baer:2013gva}, again leading to overestimates of finetuning by
factors of 10-1000\cite{Baer:2023cvi}.}
Notice that low values of $\Delta_{EW}$ require that $m_{H_u}^2$ runs barely
negative (so it has a natural value $\sim - m_Z^2$ at the weak scale) and that
$\mu$, which feeds mass to $W,\ Z,\ h$ and higgsinos, is also comparable to $m_Z$.
Thus, a prediction is that higgsinos are light, $\sim 100-350$ GeV, and
likely to result in a higgsino-like lightest SUSY particle (LSP) in SUSY models.
The mechanics of low $\Delta_{EW}$ is obvious: no implausible cancellations
between $m_{H_u}^2$, which arises from SUSY breaking but runs to very different
weak scale values, and the value of $\mu^2$, which arises from whatever solution
to the SUSY $\mu$ problem is assumed\footnote{Twenty solutions to the
  SUSY $\mu$ problem are reviewed in Ref. \cite{Bae:2019dgg}.}.
Other sparticles such as top-squarks can have masses up into the several
TeV-regime since their $\Sigma_u^u$ terms are suppressed by a loop factor.
For spectra with small $\mu\sim 100-350$ GeV,
the $\Sigma_u^u(\tst_{1,2})$ are frequently the largest contributions to
the weak scale since their Yukawa couplings are large:
\be
\Sigma_u^u(\tst_{1,2})=\frac{3}{16\pi^2}F(m_{\tst_{1,2}})\times\left[ f_t^2-g_Z^2\mp\frac{f_t^2A_t^2-8g_Z^2(\frac{1}{4}-\frac{2}{3}x_W)\Delta_t}{m_{\tst_2}^2-m_{\tst_1}^2}\right]
\label{eq:Sigmauu}
\ee
where $\Delta_t=(m_{\tst_L}^2-m_{\tst_R}^2)/2+m_Z^2\cos 2\beta(\frac{1}{4}-\frac{2}{3}x_W)$, 
$g_Z^2=(g^2+g^{\prime 2})/8$, $x_W\equiv\sin^2\theta_W$ and $F(m^2)=m^2(\log (m^2/Q^2)-1)$ with
$Q^2\simeq m_{\tst_1}m_{\tst_2}$.
Hence, top-squarks cannot be too heavy.
Since $\Sigma_u^u(\tst_{1,2})\sim m_{\tst_1}^2/16\pi^2$, the top squark masses
lie typically in the few TeV range.
(In addition, for large $A_t$ terms, there can be large cancellations
in Eq. \ref{eq:Sigmauu}.
These cancellations for both $\tst_1$ and $\tst_2$ occur for $A_t$ values
near where the light Higgs mass is maximal\cite{Baer:2012up}.)
Gluinos enter the EW scale at 2-loop order\cite{Dedes:2002dy}
but also impact naturalness by feeding into the evolution of the
squark soft terms and can range up to $\sim 6-9$ TeV
at little cost to naturalness\cite{Baer:2018hpb}.

The presence of light higgsinos leads to some very different detection
strategies for SUSY at LHC compared to earlier predictions with a
bino- or wino-like LSP\cite{Baer:2020kwz}. Dark matter is expected to occur
as a mixture of (mainly) SUSY-DFSZ-axion along with a depleted abundance of
higgsino-like WIMPs\cite{Bae:2013bva,Bae:2013hma}.

\section{Status of SUSY after LHC Run 2}
\label{sec:status}

What are we to make of the results from LHC where a very SM-like Higgs boson of
mass $m_h\simeq 125$ GeV has been discovered\cite{ATLAS:2012yve,CMS:2012zhx}, 
but as of yet no distinct
signals for SUSY have emerged? The light SUSY Higgs mass including
dominant one-loop corrections from top-quarks/squarks is given by
\be
m_h^2=m_Z^2\cos^2 2\beta +\frac{3 g^2 m_t^4}{8\pi^2m_W^2}
\left(\log (\frac{m_{SUSY}^2}{m_t^2})+ \frac{x_t^2}{m_{SUSY}^2}\left(1-\frac{x_t^2}{12 m_{SUSY}^2}\right)\right)
\ee
where $m_{SUSY}^2\simeq m_{\tst_1}m_{\tst_2}$ optimizes the scale choice and
$x_t=A_t-\mu\cot\beta$.
The value of $m_h^2$ maximizes for $x_t=\pm\sqrt{6}m_{SUSY}$,
the maximal mixing scenario\cite{Carena:2002es},
which then requires rather large weak scale trilinear soft terms. 

\begin{figure}[htb!]
\centering
    {\includegraphics[height=0.4\textheight]{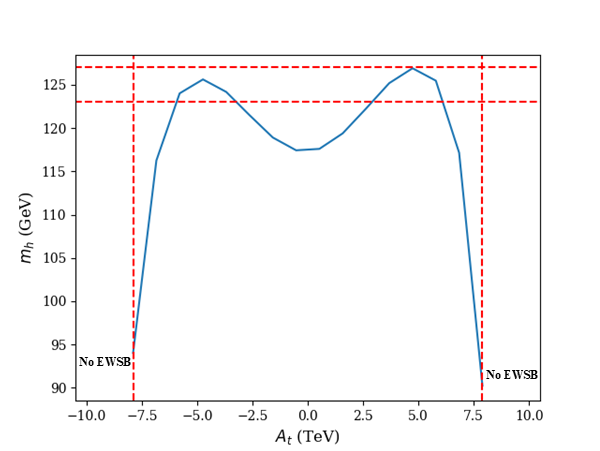}}
    \caption{Value of $m_h$ vs. $A_t(weak)$ from FeynHiggs
      for the NUHM2 model with $m_0=5$ TeV, $m_{1/2}=1.2$ TeV, $A_0=-8$ TeV
      $\tan\beta =10$, $\mu =200$ GeV and $m_A=2$ TeV.
                \label{fig:mhvsAt}}
\end{figure}

Given the paramount importance of $m_h$ in this work,
we use Isajet 7.91 to evaluate sparticle mass spectra\cite{Paige:2003mg,Baer:1994nc}, but interface with FeynHiggs\cite{Heinemeyer:1998yj}
for the most up-to-date derived values of $m_h$.
The value of $m_h$ is plotted in Fig. \ref{fig:mhvsAt} vs. $A_t(weak)$ for a
natural SUSY benchmark point in the NUHM2 model\cite{Ellis:2002wv} with $m_0=5$ TeV,
$m_{1/2}=1.2$ TeV, $A_0=-8$ TeV, $\tan\beta =10$, $\mu =200$ GeV and $m_A=2$ TeV.
The results are plotted versus $A_t(weak)$. From Fig. \ref{fig:mhvsAt}, we see that $m_h$ is minimized around $A_t\simeq 0$ (minimal mixing scenario) where $m_h\sim 116$ GeV, well below its measured value. As $|A_t|$ increases in either direction, we see $m_h$ reaching $\sim 125$ GeV for $A_t\sim m_0\sim 5$ TeV, {\it i.e.} substantial weak-scale trilinear terms are needed. For somewhat larger values of $|A_t|$, the value of $m_h$ drops precipitously
before the model enters the ``no EWSB'' region, where large values of
$A_0$ cause top-squark soft terms to run tachyonic leading to
charge-and-color-breaking (CCB) minima in the scalar potential.

\begin{figure}[htb!]
\centering
    {\includegraphics[height=0.4\textheight]{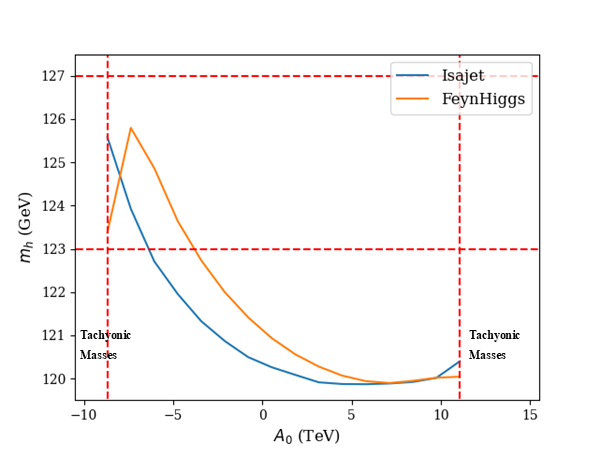}}
        {\includegraphics[height=0.4\textheight]{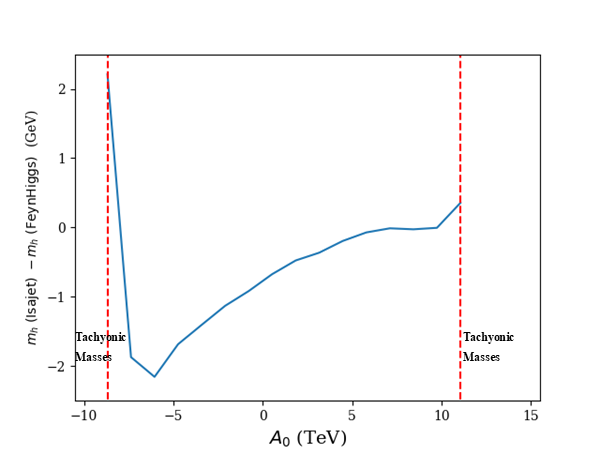}}
        \caption{In {\it a}), we show the value of $m_h$ vs. $A_0$
          from FeynHiggs and Isasugra
      for the NUHM2 model with $m_0=5$ TeV, $m_{1/2}=1.2$ TeV,
      $\tan\beta =10$, $\mu =200$ GeV and $m_A=2$ TeV.
      Frame {\it b}) shows the difference in $m_h$ between FeynHiggs and Isasugra.
                \label{fig:mhvsA0}}
\end{figure}

For reference, in Fig. \ref{fig:mhvsA0}{\it a}) we plot the value of
$m_h$ for the same NUHM2 benchmark point as in Fig. \ref{fig:mhvsAt},
but this time vs. $A_0$ using both Isajet and FeynHiggs. The value of $m_h^2$
in Isajet is computed using the RG-improved one-loop effective potential
keeping just third generation fermion/sfermion contributions\cite{Bisset:1995dc},
but where the Yukawa couplings are evaluated including weak scale
threshold corrections\cite{Pierce:1996zz} which provide the leading two-loop
effects\cite{Carena:2002es}. Frame {\it b}) shows the difference between
Isajet and FeynHiggs which in this case at least is always within $\pm 2$ GeV.
From frame {\it a}), we see the value of $m_h$ maximizing for large
{\it negative} $A_0$ values with $A_0\sim -1.6 m_0$. Using all GUT scale
parameters, the spectra hits the region of tachyonic stop masses for large
positive $A_0$ before $m_h$ reaches its maximal mixing value.

\subsection{Consequences of $m_h=125$ GeV for SUSY models with small trilinears}
\label{ssec:mh125}

The consequences of $m_h=125$ GeV for SUSY models have been known
for a long time:
\bi
\item within the MSSM, one can lift $m_h\to 125$ GeV in SUSY models with small
  trilinear soft terms $A_t\sim 0$ by requiring top-squarks in the
  $10-100$ TeV regime or
\item again within the MSSM, one can have TeV-scale top-squarks but with large $A$-terms\cite{Baer:2011ab} or
\item one may proceed beyond the MSSM and add additional fields,
  such as NMSSM singlets\cite{Hall:2011aa} or vectorlike matter fields\cite{Martin:2009bg,Martin:2010dc,Martin:2012dg}, that contribute to $m_h$.
  \ei
  In the interests of minimality, we will focus here on the first two possibilities, and avoid the question of the role of visible sector singlets in models with the
  SM gauge symmetry, or questions on why so far hints of vector-like matter
  have not yet been revealed by experiments.

  The first of these cases, for models with small $A_t$,  was shown as the
  data confirming $m_h\simeq 125$ GeV started rolling in in 2012\cite{Arbey:2011ab,Baer:2012uya,Draper:2011aa}.

  \subsubsection{Gauge mediated SUSY breaking (GMSB)}  

  In these models, a messenger sector characterized by energy scale $\Lambda$
  is posited which communicates
  between the SUSY breaking sector and the visible sector\cite{Dine:1995ag}.
  Scalars gain
  squared masses at 2-loops $m_i^2\sim \left(\frac{\alpha_i}{4\pi}\Lambda\right)^2$
  whilst gauginos gain mass at 1-loop at $m_{\lambda}\sim c\lambda \frac{\alpha_i}{4\pi}\Lambda$, so they are comparable. Trilinear soft terms are further loop suppressed so that one takes $A_i\simeq 0$ at the messenger scale $Q=\Lambda$.
  The small $A_i$ terms then require that stop masses lie in the tens of TeV
  range in order to generate $m_h\sim 123-127$ GeV.
  With such large stop masses, the $\Sigma_u^u(\tst_{1,2})$ contributions to the weak scale become of order $\sim 10^3-10^5$ and consequently the minimal
  GMSB model is highly finetuned under $\Delta_{EW}$ (see Fig. 6 of
  Ref. \cite{Baer:2014ica}).
  Alternatively, the natural value of the weak scale in mGMSB without finetuning but with  $m_h\simeq 125$ GeV is $m_{weak}\sim m_Z\sqrt{\Delta_{EW}/2}\sim 2-20$ TeV.

  \subsubsection{Gaugino mediated SUSY breaking (inoMSB)}

\begin{figure}[htb!]
\centering
    {\includegraphics[height=0.35\textheight]{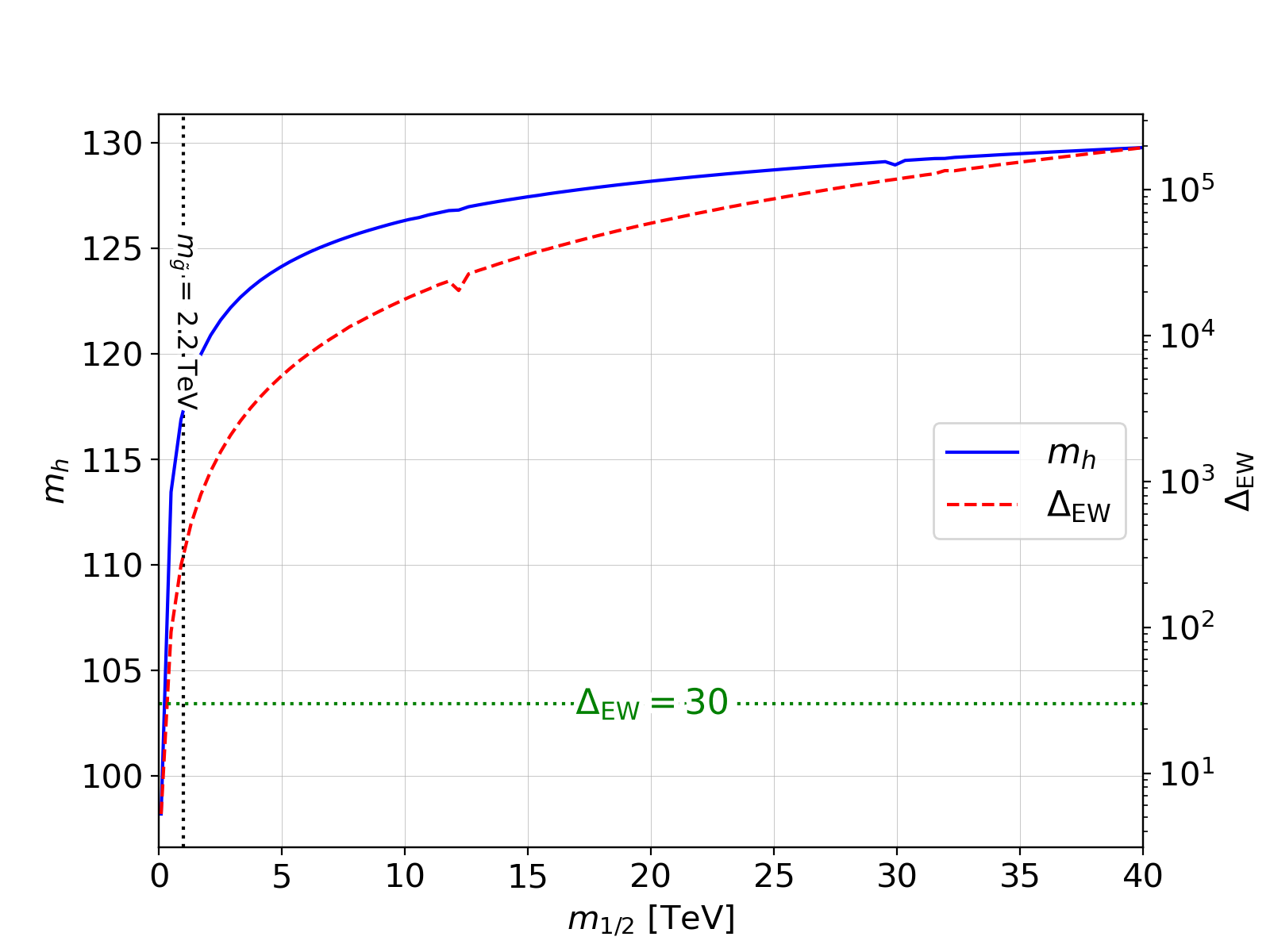}}\\
        {\includegraphics[height=0.35\textheight]{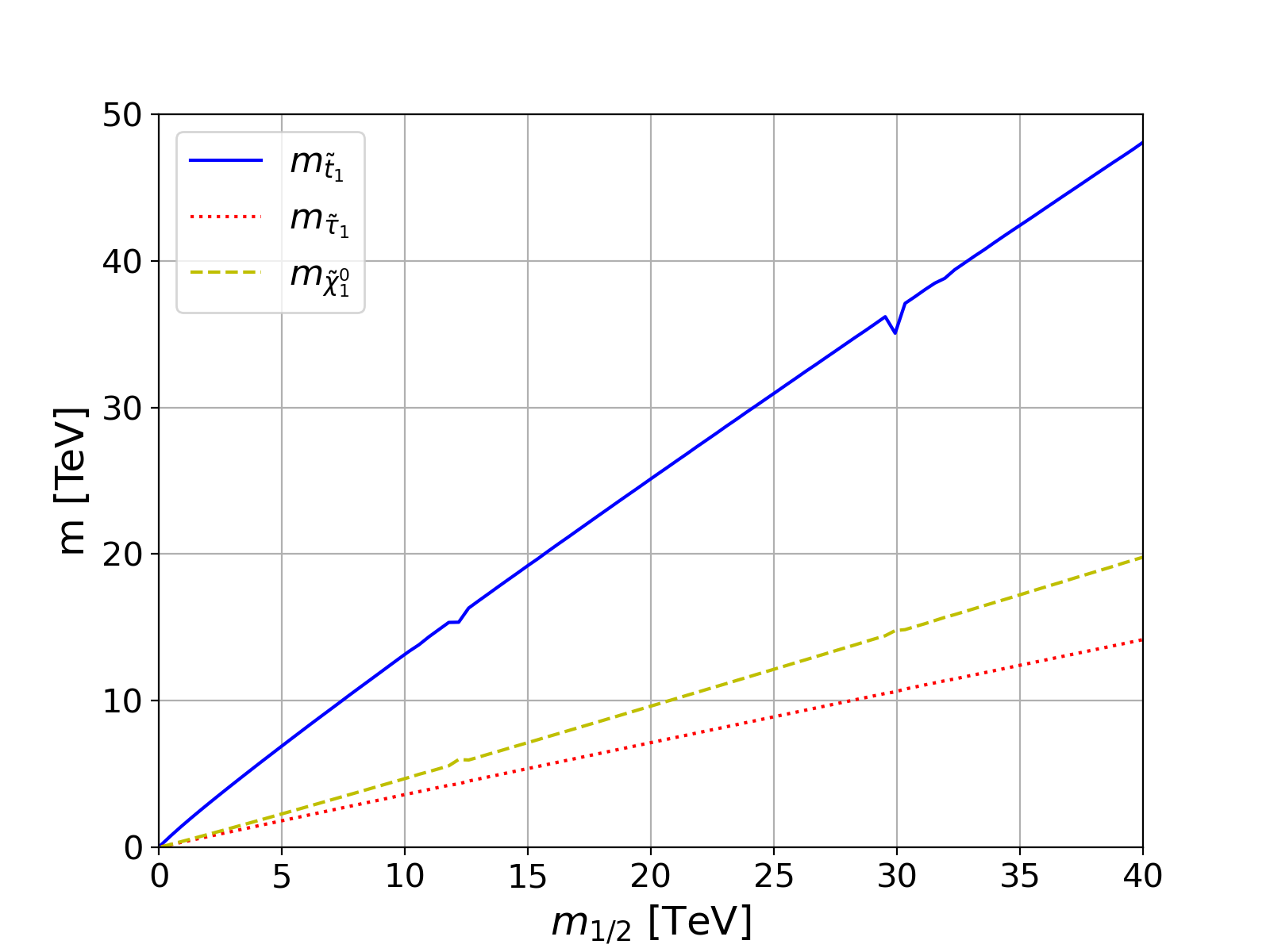}}
    \caption{Plot of {\it a}) $m_h$ (left vertical axis, blue curve) and $\Delta_{EW}$
      (right vertical axis, red curve)
      vs. $m_{1/2}$ in the inoMSB model for $\tan\beta =10$ and $\mu >0$.
In frame {\it b}), we plot several sparticle masses vs. $m_{1/2}$.
      \label{fig:inoMSB}}
\end{figure}

  In the inoMSB setup\cite{Kaplan:1999ac,Chacko:1999mi,Schmaltz:2000gy},
  matter superfields live on one brane whilst SUSY
  breaking fields live on a different brane separated from the first in an
  extra-dimensional geometry. Gravity and gauge fields live in the bulk.
  Since gauge superfields interact directly with the SUSY breaking sector,
  they gain weak scale masses $m_{1/2}$ at the compactification scale $M_c$
  whilst scalar masses and $A_i$ terms are $\sim 0$.
  The matter scalars gain mass from RG evolution from $Q=M_c$ to $Q=m_{weak}$.  
  This is reminiscent of the no-scale supergravity
  spectrum\cite{Lahanas:1986uc}. To gain $m_h\simeq 125$ GeV, enormous
  values of $m_{1/2}$ are required which drives $\Delta_{EW}$ to very large
  values: see Fig. \ref{fig:inoMSB} where we use Isasugra for the
  inoMSB spectra and FeynHiggs for the corresponding value of $m_h$.
  As is expected, in inoMSB with $m_0=0$, the lightest stau is LSP,
  so either $R$-parity violation or other LSPs (axinos, gravitinos, $\cdots$)
  would have to be invoked to avoid cosmological constraints on
  stable charged relics.

\section{Gravity-mediated SUSY breaking}

In gravity-mediated SUSY breaking models, one assumes a hidden sector
which serves as an arena for SUSY breaking, and a visible sector containing
(at least) the usual MSSM fields.
All are contained in the supergravity (SUGRA) Lagrangian, which is assumed
to be the low energy effective field theory (EFT) of
string compactification to four dimensions.
Under hidden sector SUSY breaking, then weak scale soft SUSY breaking terms
emerge from Planck-suppressed SUGRA operators.
Scalar masses arise from terms such as
\be
\int d^4\theta c_{ij}\frac{X^\dagger X Q_i^\dagger Q_j}{m_P^2}\to
c_{ij}\frac{F_X^\dagger F_X}{m_P^2}\phi_i^*\phi_j
\label{eq:mscalar}
\ee
where the $X$ fields are hidden sector SUSY breaking fields and the
$Q$ are visible sector chiral superfields and the $\theta$s are
superspace coordinates.
When the auxiliary $X$ fields gain a SUSY breaking vev $F_X$, then
scalar soft squared masses $m_{ij}^2\sim c_{ij}\frac{F_X^\dagger F_X}{m_P^2}$
arise and if $F_X\sim (10^{11}$ GeV)$^2$, then weak scale
soft scalar mass terms will ensue.
In the above, we leave the indices $i,j$ as arbitrary
(so long as $Q_i^\dagger Q_j$ is invariant under the assumed
(gauge and possibly other) symmetries to emphasize that nonuniversal
scalar masses\cite{Soni:1983rm,Kaplunovsky:1993rd,Brignole:1993dj}
are the rule and not the exception in SUGRA.
In this sense, models like CMSSM/mSUGRA
with an assumed ad-hoc scalar mass universality are not likely a realistic
expression of gravity-mediation. Also, since the scalar masses arise from
$F_X^\dagger F_X$, they are not protected by any hidden sector symmetries
which might enforce small (weak scale) values.

Gaugino masses in SUGRA arise via
\be
\int d^2\theta \frac{Y W^\alpha W_\alpha}{m_P}\to \frac{F_Y}{m_P}\lambda\lambda
\label{eq:mino}
\ee
leading to weak scale gaugino masses when the hidden sector field $Y$
attains a SUSY breaking vev $F_Y$. It is important to notice here
that the SUSY breaking field $Y$ must be a gauge singlet and further that
generation of gaugino masses leads to a broken $R$-symmetry in
the Lagrangian since the gauginos carry $R$-charge $+1$.

Trilinear terms arise as
\be
\int d^2\theta \frac{Y H_uQ U^c}{m_P}\to \frac{F_Y}{m_P} H_u\tilde{Q}\tilde{u}_R
\label{eq:Aterms}
\ee
   {\it etc.} so that trilinear soft terms $a_{ijk}\sim m_{weak}$ also
   develop.
   As for gaugino masses, the field $Y$ must be a hidden sector singlet.

   The SUSY conserving $\mu$ term would be expected to arise with $\mu\sim m_P$
   whereas phenomenology requires $\mu\sim m_{weak}$.
   This is the SUSY $\mu$ problem\cite{Bae:2019dgg}.
   First, $\mu$ must be forbidden
   by some mechanism/symmetry, and then regenerated at the weak scale.
   We prefer including the SUSY DFSZ axion solution where the global $U(1)_{PQ}$
   arises accidentally and approximately from a discrete $R$-symetry such as
   ${\bf Z}_{24}^R$\cite{Lee:2011dya,Baer:2018avn,Bhattiprolu:2021rrj}.
   This approach generates $\mu\sim m_{weak}$ via the
   Kim-Nilles\cite{Kim:1983dt} superpotential operator $W\ni \lambda_\mu S^2H_uH_d/m_P$
   thus solving the strong CP problem while generating both PQ and $R$-parity symmetries. The discrete $R$-symmetry is strong enough to solve the axion quality problem as well.

Bilinear soft terms come from the operator
   \be
   \int d^4\theta X^\dagger X H_u H_d/m_P^2\to B_\mu\sim (F_X/m_P)^2 
   \ee
and are also of order the weak scale.
   
   For gravity-mediated SUSY breaking, we consider two cases:
\begin{enumerate}
\item the case where SUSY breaking fields carry some hidden sector charge and
\item singlet SUSY breaking fields in the hidden sector.
\end{enumerate}

\subsection{Charged SUSY breaking fields}

Charged SUSY breaking is motivated by early renditions of dynamical
SUSY breaking\cite{Witten:1981nf} wherein SUSY breaking occurs
non-perturbatively in the hidden sector perhaps via
gaugino condensation\cite{Ferrara:1982qs} $\langle \lambda\lambda\rangle$
where gauginos end up condensing
at a scale $\Lambda_{GC}$ where the hidden sector gauge coupling $g_{hidden}$
becomes strongly interacting.
These scales are related  by dimensional transmutation to the
scale of hidden sector SUSY breaking via
$m_{hidden}^2\sim \Lambda_{GC}^3/m_P\sim m_P^2 \exp^{-8\pi^2/g_{hidden}^2}$
where $g_{hidden}$ is the asymptotically free hidden gauge sector coupling.
In such early models, singlet SUSY breaking fields were scarce or non-existent
leading to a gaugino mass problem\cite{Affleck:1984xz}.
This situation motivated the $4-d$ introduction of what became known as
anomaly-mediated SUSY breaking\cite{Giudice:1998xp} (AMSB) wherein
matter and Higgs scalars could gain large masses but where gauginos
gained loop-suppressed masses proportional to the gauge group beta
functions\cite{Kaplunovsky:1994fg}. Unlike the extra-dimensional sequestering
mechanism of Ref. \cite{Randall:1998uk}, the $4-d$ DSB-inspired version of
AMSB has no problem with tachyonic slepton masses and no need for ad-hoc bulk
contributions to soft breaking terms since large scalar masses are expected to be generically 
present.

Charged SUSY breaking\cite{Giudice:1998xp} led to what became known as PeV-SUSY\cite{Wells:2004di}
and then mini-split\cite{Arvanitaki:2012ps} SUSY;
it was emphasized by Wells\cite{Wells:2003tf} that a loop-level
hierarchy between gauginos and $A$-terms as compared to scalar soft masses could ensue.
This may have served as an inspiration for split SUSY\cite{Arkani-Hamed:2004ymt,Giudice:2004tc,Arkani-Hamed:2004zhs} where
the gaugino-to-scalar mass hierarchy was taken to be far larger.
Split SUSY predicted the light Higgs to have mass in the range
$m_h\sim 140-160$ GeV\cite{Giudice:2011cg} so that once the Higgs was
discovered with $m_h\simeq 125$ GeV, the need became apparent to scale
back the scalar masses to the 10-100 TeV range whch could generate
$m_h\sim 125$ GeV even with small $A$-terms.

Charged SUSY breaking (CSB) is well described by the preprogrammed mAMSB
model embedded in Isajet\cite{Paige:2003mg}.
The CSB model has parameter space
\be
m_0,\ m_{3/2},\ \tan\beta,\ sign(\mu )\ \ \ \ (CSB)
\ee
where the pure AMSB contributions to scalar masses, $A$-terms and gaugino
masses are included. Thus, we expect the loop-suppressed AMSB form for
gaugino masses and $A$-terms but scalar masses can be much larger and
are at first approximated by a universal value $m_0$.
The value of $B_\mu$ is traded for $\tan\beta$ using the EWSB
minimization conditions.

\begin{figure}[htb!]
\centering
    {\includegraphics[height=0.35\textheight]{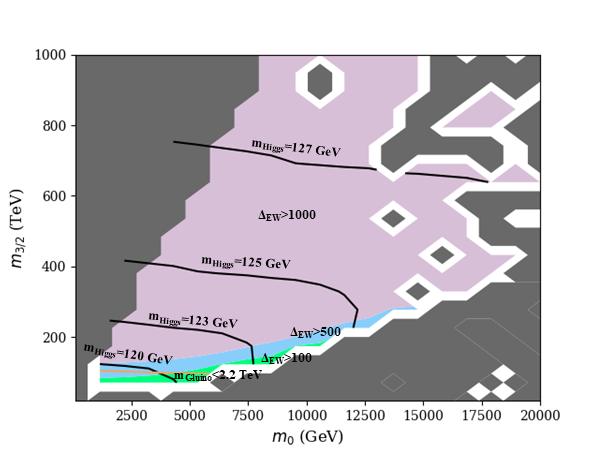}}\\
        {\includegraphics[height=0.35\textheight]{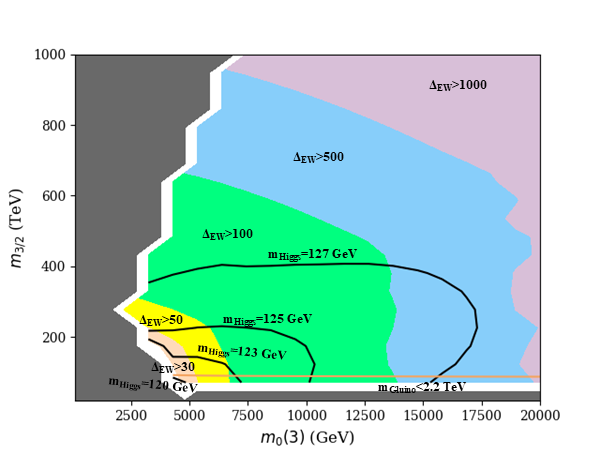}}
    \caption{Parameter space of  
      charged SUSY breaking (CSB) model with {\it a})
      $\tan\beta =10$ in the $m_0$ vs. $m_{3/2}$ plane.
      In {\it b}), we show CSB parameter space with non-universal bulk terms
      such that $\mu =200$ GeV and $m_A=2$ TeV but still with $A_0$
      as given by its loop-suppressed AMSB value.
      \label{fig:mamsb}}
\end{figure}

The $m_0$ vs. $m_{3/2}$ parameter space plane for charged SUSY breaking is
shown in Fig. \ref{fig:mamsb}{\it a}) for a fixed value of $\tan\beta =10$
and $\mu >0$.
The left-most gray area is excluded because it gives the well known
tachyonic sleptons.
The lower-right gray region has inappropriate EWSB.
The upper-right disconnected gray regions indicate an instability of the code in
calculating the sparticle mass spectra with huge scalar masses.
The region below the brown contour is excluded by LHC
limits that $m_{\tg}\agt 2.2$ TeV.
(The unshaded regions around the gray areas are an artifact of the graphing
software used.)
We also show contours of $m_h$ as computed by FeynHiggs interfaced with
Isasugra.
Assuming a $\sim \pm 2$ GeV theory error on $m_h$,
then the region between $123$ GeV$<m_h<127$ GeV would be
allowed by the LHC Higgs mass measurement.
With $A_0$ small, the lower-left region favored by early
phenomenological analyses is then excluded.
The color-shaded regions correspond to different ranges of $\Delta_{EW}$:
orange has $\Delta_{EW}>30$, yellow $\Delta_{EW}>50$, green $\Delta_{EW}>100$,
blue $\Delta_{EW}>500$, and purple $\Delta_{EW}>1000$.
From the plot, we see that the $m_h$-allowed region mainly has $\Delta_{EW}>1000$, aside from a small sliver just beyond the $m_{\tg}\sim 2.2$ contour
where $\Delta_{EW}\sim 100-1000$.
Without finetuning, this region would correspond to a derived weak scale in
the vicinity of $m_{weak}\agt 2$ TeV, exemplifying the Little Hierarchy Problem.

In Fig. \ref{fig:mamsb}{\it b}), we adopt non-universal scalar masses
with $m_{H_u}\ne m_{H_d}\ne m_0$ so that $\mu$ can be set to a natural
value $\mu =200$ GeV.
Here, the naturalness problem is improved, but not eliminated,
since $\Delta_{EW}$ ranges from $40-700$ in the region where
$m_h\sim 123-127$ GeV.
Even with a natural value of $\mu$, the small AMSB $A_0$-term necessitates
unnatural top-squark contributions to the weak scale in order to gain
$m_h\simeq 125$ GeV in accord with LHC measurements.
This corresponds to a predicted weak scale, without fine-tuning,
of $m_Z\sim 400-1700$ GeV.

\subsection{Singlet SUSY breaking fields}

Singlet superfields in older models of DSB seemed to be scarce\cite{Affleck:1984xz} leading to a gaugino mass problem: how to generate visible sector
gaugino masses which were comparable to the other soft terms.
Nelson had shown a path forward\cite{Nelson:1995hf}, modifying early DSB models
so that  gauge singlets did get $F$ terms of order $F_X\sim m_{3/2}m_P$, and then
MSSM gauginos could gain mass values comparable to the soft scalar masses.
New possibilities were opened with the advent of metastable DSB
(MDSB)\cite{Intriligator:2006dd,Intriligator:2007py} where the SUSY breaking minimum was in fact not the global minimum of the theory.
The new MDSB models seem generic in field theory and string theory and rather easy to implement.
Dine {\it et al.}\cite{Dine:2006gm} showed that by the process of retrofitting, one could adopt the simple
spontaneous SUSY breaking models and replace the mass terms, naively expected to be of order $m_P$, by a strongly interacting hidden gauge sector where the
mass scale was instead exponentially suppressed by non-perturbative effects
such as gaugino condensation.
This could work in gauge mediation\cite{Dine:2006gm} or gravity mediation\cite{Bose:2012gq}.

As an example of retrofitting in gravity mediation, we assume the Polonyi
superpotential\cite{Polonyi:1977pj} with a single gauge singlet hidden sector
superfield $X$:
\be
W_{Polonyi}= m^2m_P(X/m_P+\beta )
\ee
where $m^2\sim m_{3/2}m_P$ and $\beta$ is dimensionless and $\sim 1$.
The intermediate scale $m\sim 10^{11}$ GeV is needed to generate
$m_{soft}\sim m^2/m_P\sim m_{3/2}\sim m_{weak}$, but is awkward to realize in
SUGRA where the only mass parameter is the Planck mass $m_P$.

In retrofitting\cite{Bose:2012gq}, we replace $m^2\to W_\alpha W^\alpha/m_P$
where $W_\alpha$ is a hidden sector gauge superfield transforming under
hidden sector group $SU(N)$ with coupling $g_H$.
We can assign an $R$-charge of $+1$ for the gauge superfields and zero for the
$X$ superfield.
We expect gaugino condensation when the coupling $g_H$ becomes strong at a scale
$\langle\lambda\lambda\rangle\sim \Lambda^3\sim m_{3/2}m_P^2\sim m_P^3\exp (-8\pi^2/g_H^2)$ by dimensional transmutation leading to breaking of an
$R$-symmetry of the hidden gauge fields but not SUSY.
But then, as a consequence, SUSY is broken by the auxiliary field $F_X$ of
$X$ gaining a vev $F_X\sim m_{3/2}m_P$ and $\langle X\rangle\sim m_P$
via the Polonyi superpotential\cite{Baer:2006rs}. 

\begin{figure}[htb!]
\centering
    {\includegraphics[height=0.4\textheight]{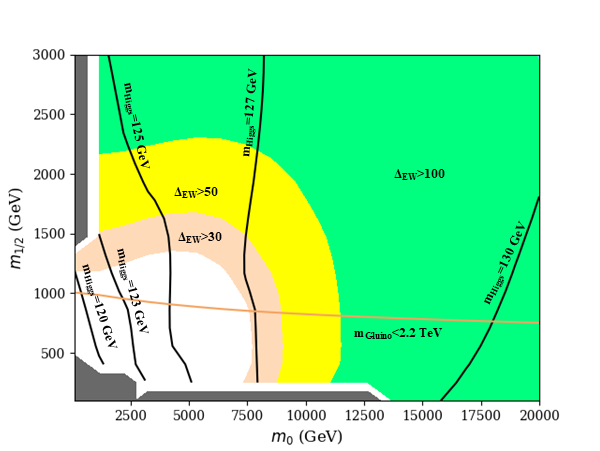}}
    \caption{Parameter space of the singlet sector gravity-mediation model
      NUHM2 in the $m_0$ vs. $m_{1/2}$
      plane for $A_0=-1.6 m_0$, $\tan\beta =10$, $\mu =200$ GeV
      and $m_A=2$ TeV. 
      \label{fig:nuhm2}}
\end{figure}

In Fig. \ref{fig:nuhm2}, we show the $m_0$ vs. $m_{1/2}$ plane of the
two-extra-parameter non-universal Higgs model with parameter space
given by\cite{Ellis:2002wv,Baer:2005bu}
\be
m_0,\ m_{H_u},\ m_{H_d},\ m_{1/2},\ A_0,\ \tan\beta,\ sign(\mu )\ \ \ \text{(NUHM2)}
\ee
which should be the expected expression of gravity-mediation models
with singlets in the hidden sector\cite{Soni:1983rm,Kaplunovsky:1993rd,Brignole:1993dj}.
The various scalars in different gauge multiplets have
non-universal masses, as expected, although this case (for simplicity)
takes the different generations of scalars to be unrealistically degenerate.
The phenomenology will roughly be the same unless the scalars drift into the
multi-TeV region.
(The case of decoupled quasi-degenerate first/second generation scalars
is shown in Ref. \cite{Baer:2024hpl}; this latter case offers a landscape
quasi-degeneracy/decoupling solution to the SUGRA flavor and CP
problems\cite{Baer:2019zfl}.)
It is convenient to trade the two soft breaking Higgs masses for the
weak scale values of $\mu$ and $m_A$ via the scalar potential minimization
conditions\cite{Ellis:2002wv}.
The plot Fig. \ref{fig:nuhm2} is generated with Isasugra interfaced with FeynHiggs.

Since we expect hidden sector singlets, we also expect gaugino masses
$m_{1/2}\sim m_0$ and large $A$-terms. In Fig. \ref{fig:nuhm2},
we take $A_0=-1.6 m_0$ and $\tan\beta =10$ with $\mu =200$ GeV and $m_A=2$ TeV.
The shaded regions are similar to Fig. \ref{fig:mamsb}.
From the plot, we see a broad unshaded region in the lower left
with $\Delta_{EW}<30$, corresponding to a derived weak scale (without tuning)
comparable to our measured value $m_{weak}^{OU}\sim m_{W,Z,h}\sim 100$ GeV.
A large bulge of allowed parameter space actually stands out beyond LHC
gluino mass limits.
This higher $m_{1/2}$ natural region is actually favored by rather general
considerations of the string landcape,
where a power-law draw\cite{Douglas:2004qg,Susskind:2004uv,Arkani-Hamed:2005zuc}
to large soft terms is exected so long as the derived value of the
weak scale lies within the ABDS window\cite{Agrawal:1997gf} $m_Z^{OU}/2\alt m_Z^{PU}\alt (2-5)m_Z^{OU}$ (where $m_Z^{PU}$
is the pocket-universe value of $m_Z$ while $m_Z^{OU}$ is the $Z$ mass
in our universe). Thus, the allowed high $m_{1/2}$ natural region is
actually favored by what Douglas\cite{Douglas:2004zg} calls
{\it stringy naturalness}\cite{Baer:2019cae}. 
The stringy natural region largely has $123$ GeV$<m_h<127$ GeV due to the
large $A$-terms generated by the presence of hidden sector
SUSY breaking gauge singlets.

It is sometimes stated in the literature that landscape selection of
parameters offers an alternative to naturalness.
After all, landscape selection could lead to the highly finetuned value of
the cosmological constant\cite{Weinberg:1987dv} $\Lambda_{CC}$.
However, in Weinberg's scheme, one can argue that the observed value of
$\Lambda_{CC}$ is about as natural as possible given the anthropic
requirement that the universe not expand so fast that galaxies fail to condense.
In the case of SUSY, if models with a natural value of the weak scale are
possible, then they will be preferred statistically over unnatural models.
This is bourne out in Ref. \cite{Baer:2022wxe} where natural SUSY models--
with all contributions to the weak scale being comparable to the weak scale--
have a far larger volume of parameter space on the landscape than
finetuned models.
The available parameter space of finetuned models shrinks to a tiny volume
(after all, they require finely-tuned parameters to gain intricate
cancellations such that the weak scale remains around 100 GeV).
In fact, in the case of tuning the $\mu$ parameter, the relative probabilities
of different models can be computed on the landscape.
For instance, in Ref. \cite{Baer:2022dfc} it is found that certain finetuned
models of mini-split or PeV-SUSY occur at probabilities of $\sim 10^{-4}-10^{-8}$
with respect to natural models, based on the volume of parameter space on
the landscape.

To summarize, the measured mass of the Higgs boson $m_h\simeq 125$ GeV
is most plausibly realized in models with large trilinear soft terms,
which then points to gravity-mediated SUSY breaking models with hidden sector
gauge singlets. While these latter fields seemed problematic in early
renditions of DSB, they seem much more generic in MDSB and retrofitted
SUGRA models. They also point to gaugino masses comparable to scalar
masses, as opposed to a little hierarchy between gauginos and scalars
as would be expected in minisplit.

\section{Digression on hidden sector effects on visible sector soft term running}

So far, our analysis has made the implicit assumption that the MSSM is the
LE-EFT valid between energy scale $Q=m_{GUT}$ and $Q=m_{weak}$.
This assumption is supported in part by the fact that gauge couplings unify
under MSSM running and electroweak symmetry is radiatively broken by
renormalization group running effects with a large top-quark mass $m_t\sim 173$
GeV. However, the situation of hidden sector effects on visible sector
soft term running has been explored in several papers starting with
Luty and Sundrum\cite{Luty:2001zv,Dine:2004dv,Cohen:2006qc,Murayama:2007ge,Craig:2009rk}.
In these papers, the example is given of a superconformal strongly
interacting hidden sector interacting with the visible sector which affects
visible sector soft term running between the scales $Q=m_{GUT}$ and
an intermediate scale $Q\sim m_{int}$ where the hidden sector is integrated out.
In particular, the hidden sector running can cause a power-law suppression of
unwanted operators and thus aiding in some model building problems.
Examples include solving the $B\mu$ problem in gauge mediation and providing
suppression/sequestering of scalar masses in models like CSB mentioned above,
which may aid in solving the SUSY flavor problem.
In the scalar sequestering model, strong superconformal running at high scales
can crunch the scalar masses towards zero at the intermediate scale.
And assuming the $\mu$ term arises via the Giudice-Masiero (GM)
mechanism (not KN as assumed herein) then the combinations $\mu^2+m_{H_{u,d}}^2$
are driven to zero at the intermediate scale. The latter effect introduces
a correlation such that $\mu^2\simeq -m_{H_{u,d}}^2$ at the intermediate scale
so that these terms are no longer independent as assumed in Eq. \ref{eq:mzs}.

The phenomenological consequences of strong scalar sequestering
(where hidden sector running dominates MSSM running for $Q>m_{int}$)
have been examined in Ref's \cite{Perez:2008ng,Martin:2017vlf,Baer:2024zvr}.
Here, the boundary conditions are that sfermion soft masses and the combinations
$\mu^2+m_{H_{u,d}}^2\to 0$ at a scale $m_{int}\sim 10^{11}$ GeV, and where MSSM
running ensues at lower energy scales $Q<m_{int}$. Examination of the spectra
of these models with strong scalar sequestering finds that over the bulk of
parameter space either no or improper EWSB occurs, and in regions where
proper EWSB occurs, then the lightest SUSY particle (LSP) is charged
(usually the right-slepton), in violation of cosmology which forbids
stable charged relics from the Big Bang. This situation can be remedied by
imposing R-parity violation so that the LSP decays before BBN.

Martin\cite{Martin:2017vlf} has identified an in-between case of moderate
scalar sequestering due to upper limits on anomalous dimensions which give
rise to comparable hidden sector/MSSM running at $Q>m_{int}$.
In this case, scalar masses and $\mu^2+m_{H_{u,d}}^2$ run to fixed-points at
$Q=m_{int}$ rather than to zero.
This improves the phenomenology, although for unified gaugino masses,
the LSP still ends up being charged.
For non-unified gaugino masses, an acceptable phenomenology can be
found\cite{Martin:2017vlf,Baer:2024zvr}.
Also, in this case, a modified naturalness measure
$\Delta_{EW}^\prime$\cite{Baer:2024zvr} is needed which combines
the now-dependent quantities $\mu^2$ and $m_{H_{u,d}}^2$ into a single entity.
Even using $\Delta_{EW}^\prime$, our concern with finetuned models
with $m_h=125$ GeV and small $A$-terms still holds since these require
10-100 TeV top-squarks and $\Sigma_u^u(\tst_{1,2})$ still gives a huge
contribution to the weak scale which necessitates EW finetuning.\footnote{
And the above discussion illustrates an advantage of $\Delta_{EW}$ and
$\Delta_{EW}^\prime$ over other finetuning measures in that the former
are independent of any unknown high scale effects and depend only on the weak
scale spectra, and not on how it comes about.}
In light of these considerations, we feel that a neglect of hidden sector
effects on visible sector running is a warranted assumption for our analysis.

\section{Conclusions}
\label{sec:conclude}

The current results from LHC Run2, where a very SM-like Higgs boson of mass
$m_h=125$ GeV has been discovered with (as yet) no sign of supersymmetry,
can be seen as greatly restricting expectations for hidden sector
supersymmetry breaking. SUSY models with small weak scale trilinear
soft terms $A_t$ such as GMSB and inoMSB require scalar masses in the
10-100 TeV range in order to generate $m_h\sim 125$ GeV.
But such heavy third generation scalars provide huge contributions to
the weak scale rendering such models unnatural and hence implausible.
They point to gravity-mediation as the more likely mechanism to communicate
SUSY breaking from the hidden sector to the visible sector.
But gravity mediation can arise in two different guises: charged or singlet
hidden sector SUSY breaking fields. Under charged SUSY breaking, one expects
a loop level suppression of both $A$-terms and gaugino masses leading to
PeV-scale SUSY also known as mini-split. The small $A$-terms again require
third generation scalars $\sim 10-100$ TeV leading to large unnatural
values of the weak scale. The unnaturalness cannot be hidden by appealing
to landscape reasoning since such unnatural models have tiny volumes of
landscape parameter space compared to natural models.
Thus, we expect that the LHC results of $m_h\simeq 125$ GeV with as yet
no sign of SUSY point to hidden sector singlets involved in SUSY breaking
which can yield models such as NUHMi ($i=2-4$) which contain large portions of
parameter space where the Little Hierarchy Problem is not there.
Gauge singlet SUSY breaking fields can be found more commonly in MDSB
and retrofitted models than earlier renditions of DSB.

{\it Acknowledgements:} 

%This material is based upon work supported by the U.S. Department of
%Energy, Office of Science, Office of High Energy Physics under Award
%Number DE-SC-0009956.
We thank M. Dine for an email.
VB gratefully acknowledges support from the William F. Vilas estate.
HB gratefully acknowledges support from the Avenir Foundation.

%%%%%%%%%%%%%%%%%%%%%%%%%%%%%%%%%%%%%%%%%%%%%%%%%%%%%%

%\section*{References}
\bibliography{hidden}
\bibliographystyle{elsarticle-num}

\end{document}